\documentclass[sigconf]{nimeart}

\usepackage{xurl}
\usepackage{listings}
\usepackage{graphicx}
\usepackage{tikz}
\usetikzlibrary{arrows.meta, calc, decorations.pathreplacing, patterns}
\usepackage{caption}
\usepackage{longtable}
\usepackage{array}
\usepackage{enumitem}

\definecolor{codegreen}{rgb}{0.25,0.50,0.35}
\definecolor{codegray}{rgb}{0.5,0.5,0.5}
\definecolor{codeblue}{rgb}{0.15,0.30,0.55}
\definecolor{backcolour}{rgb}{0.97,0.97,0.97}
\lstdefinelanguage{json}{
  basicstyle=\footnotesize\ttfamily,
  stringstyle=\color{codegreen},
  morestring=[b]",
  literate=
    *{:}{{{\color{codeblue}:}}}{1}
    {,}{{{\color{codegray},}}}{1}
    {\{}{{{\color{black}\{}}}{1}
    {\}}{{{\color{black}\}}}}{1}
    {[}{{{\color{black}[}}}{1}
    {]}{{{\color{black}]}}}{1},
}

\AtEndPreamble{\hypersetup{colorlinks=true, linkcolor=black, citecolor=black, urlcolor=black, filecolor=black}}

\AtBeginDocument{%
  }

\setcopyright{cc}
\copyrightyear{2026}
\acmYear{2026}
\acmDOI{}

\acmConference[NIME '26]{International Conference on New Interfaces for Musical Expression}{June 23--26, 2026}{London, UK}
\acmISBN{}

\begin{document}

\title{A Text-Steerable Instrument for Sketching Procedural Soundscapes via Language Models}

\author{Prabal Gupta}
\affiliation{%
  \institution{Rama Labs}
  \city{Kitchener}
  \state{Ontario}
  \country{Canada}}
\email{prabal@rjeinc.ca}

\renewcommand{\shortauthors}{Gupta}

\begin{abstract}
We present a real-time musical interface that converts natural-language scene descriptions into evolving procedural soundscapes. A performer types a prompt such as ``warm jazz caf\'{e} at midnight'' and steers it through direct parameter adjustments---stepping brightness down, switching a rhythm style---each producing a predictable, audible shift without re-prompting. Where GPU-bound text-to-audio systems synthesize monolithic waveforms, our instrument generates human-readable configurations over a categorical schema, enabling fine-grained performer control; most valid combinations are designed to sound musically coherent. Three interchangeable backends---embedding retrieval for sub-second CPU-only use, hosted LLMs via API, and a fine-tuned 270M local model---all emit the same schema. A live generator architecture continuously emits audio while resolving new instructions in the background, crossfading seamlessly when ready; even when an LLM takes 5--12 seconds to respond, the audience hears uninterrupted sound---reframing text-to-music as an ongoing performable stream rather than a one-shot generation. We evaluate text--audio semantic alignment using LAION-CLAP on held-out prompts as a technical proxy, finding that retrieval-based configuration outperforms random valid configurations on this metric, while noting that LAION-CLAP also informed retrieval-map construction. We report performance observations, informal listener feedback, and release materials for the SDK, dataset artifacts, model, and audiovisual performance interface.
\end{abstract}

\keywords{text-to-music, procedural synthesis, real-time performance, language models, parametric control, live coding}

\maketitle

\section{Introduction}

In a live coding set, the performer types \texttt{"neon city at 2am"} and the ambient drone keeps playing. Five seconds later, the new configuration resolves---bright, electronic, fast---and the sound crossfades into it. The performer nudges brightness down two steps. The city gets darker. They switch rhythm to \texttt{"heartbeat"}. The pulse slows. None of this required stopping the music, waiting for a GPU, or understanding DSP.

Neural text-to-audio systems such as MusicGen~\cite{copet2023musicgen}, MusicLM~\cite{agostinelli2023musiclm}, and Stable Audio~\cite{evans2024stableaudio} produce high-quality audio from text. However, such systems typically return generated audio rather than interpretable synthesizer parameters; CTAG explicitly frames this as a tweakability problem for neural audio systems~\cite{cherep2024ctag}. Azimi and Zareei~\cite{azimi2025live} fine-tuned Stable Audio Open for live improvisation, demonstrating that even with GPU acceleration, generation latency and unpredictable outputs remain challenges.

Our SDK addresses this tension by generating \emph{interpretable parameter configurations} for a procedural synthesizer rather than raw audio. Its live generator architecture decouples instruction resolution from audio emission---the current configuration keeps sounding while the next one resolves in the background. This brings LLM-powered text-to-music into live performance without audible interruption.

For the performer, the key affordance is persistent steerability: a prompt establishes a scene, but the musical work happens through named parameter edits---darkening the spectrum, switching rhythm, widening the space---while the stream continues. Text becomes a playable control surface, not a one-shot trigger.

The system is not a neural audio generator: language models resolve prompts into symbolic synthesizer configurations, and a deterministic procedural engine renders the sound---less timbrally general than neural text-to-audio, but inspectable, steerable, and stable during performance.

We contribute: (1)~a live generator architecture that hides resolution latency behind continuous audio, enabling the SDK to function as an instrument runtime rather than a generation endpoint; (2)~a 34-field configuration schema whose categorical labels mean most valid combinations sound musically coherent---a mapping-design contribution in the sense of Hunt and Kirk~\cite{hunt2000mapping} and Magnusson~\cite{magnusson2019sonic}; and (3)~an open-source release.\footnote{SDK and demo: \url{https://github.com/prabal-rje/latentscore}, \url{https://latentscore.com}. Supplementary materials: \url{https://zenodo.org/records/19944277}.}
A companion ACM SIGGRAPH 2026 talk addresses the broader system architecture and benchmark framing; the present paper focuses on the instrument contribution: the live generator as performance runtime, the configuration schema as mapping design, and the prompt--listen--steer workflow.

\section{Related Work}

\paragraph{Text-to-audio instruments.}
Azimi and Zareei~\cite{azimi2025live} fine-tuned Stable Audio Open for improvisation, generating raw audio fragments. Shepardson et~al.~\cite{shepardson2024tungnaa} used text notation to control neural vocal synthesis in Tungn\'{a}\'{a}. Both systems synthesize neural audio directly, with differing compute requirements and limited post-generation editability. Our SDK generates structured parameters instead---trading timbral fidelity for sub-second latency, direct editability, and reproducible output.

\paragraph{Text-to-synthesizer mapping.}
CTAG~\cite{cherep2024ctag} maps text to a modular synthesizer with 78~parameters via iterative optimization, but its search runs at generation time. SynthScribe~\cite{brade2024synthscribe} supports synthesizer sound retrieval, creation, and modification through multimodal text/audio tools. Our system front-loads LLM generation into dataset construction and retrieves complete multi-layer configurations at runtime. Latent-space instruments like Stacco~\cite{privato2024stacco} and Latent Mappings~\cite{murraybrowne2021latent} navigate continuous neural spaces through physical gesture; our SDK navigates a discrete parameter space through language.

\paragraph{Programming frameworks.}
ChAI~\cite{li2024chai} and ChuMP~\cite{shaheed2025chump} added interactive AI tools and modular package management to ChucK.

\section{The Instrument}

\subsection{Design Goals}
\vspace{-2pt}

\emph{Immediacy~(G1):} sub-second text-to-sound. \emph{Controllability~(G2):} all parameters exposed as named fields with human-readable labels. \emph{Reproducibility~(G3):} a given configuration and seed yield identical output. \emph{Learnability~(G4):} a single-line entry point that scales to full programmatic control. \emph{Continuity~(G5):} audio never stops during configuration resolution.

\vspace{-2pt}
\subsection{The Live Generator}

The SDK's core contribution is the live generator---an async streaming interface where the instrument is always sounding. A performer (or game engine, installation, or MIDI controller) writes a Python generator that yields instructions: text prompts, absolute configurations, or relative parameter adjustments. The SDK continuously emits audio chunks from the current configuration. When a new instruction arrives, it resolves in the background---calling an LLM, querying the retrieval index, or applying a parameter update---while audio emission continues uninterrupted. Once resolved, the SDK crossfades into the new configuration.

\begin{figure}[t]
\begin{lstlisting}
async def performance():
    yield "warm jazz cafe at midnight"  # text -> config
    await asyncio.sleep(8)              # play for 8s

    yield MusicConfigUpdate(            # relative nudge
        brightness=Step(-2),            # two steps darker
        echo="heavy")                   # absolute value
    await asyncio.sleep(8)

    yield "neon rain on empty streets"  # new scene
    # current config keeps playing while backend resolves...

session = live(performance())
session.play(seconds=60)
\end{lstlisting}
\caption{The live generator programming model. The performer yields instructions; audio plays continuously from the current configuration.}
\Description{Python code listing showing an async generator function that yields text prompts and MusicConfigUpdate objects with sleep calls between them, followed by two lines creating a live session and playing for 60 seconds.}
\label{fig:code}
\end{figure}

\begin{figure*}[t]
  \centering
  \includegraphics[width=\textwidth]{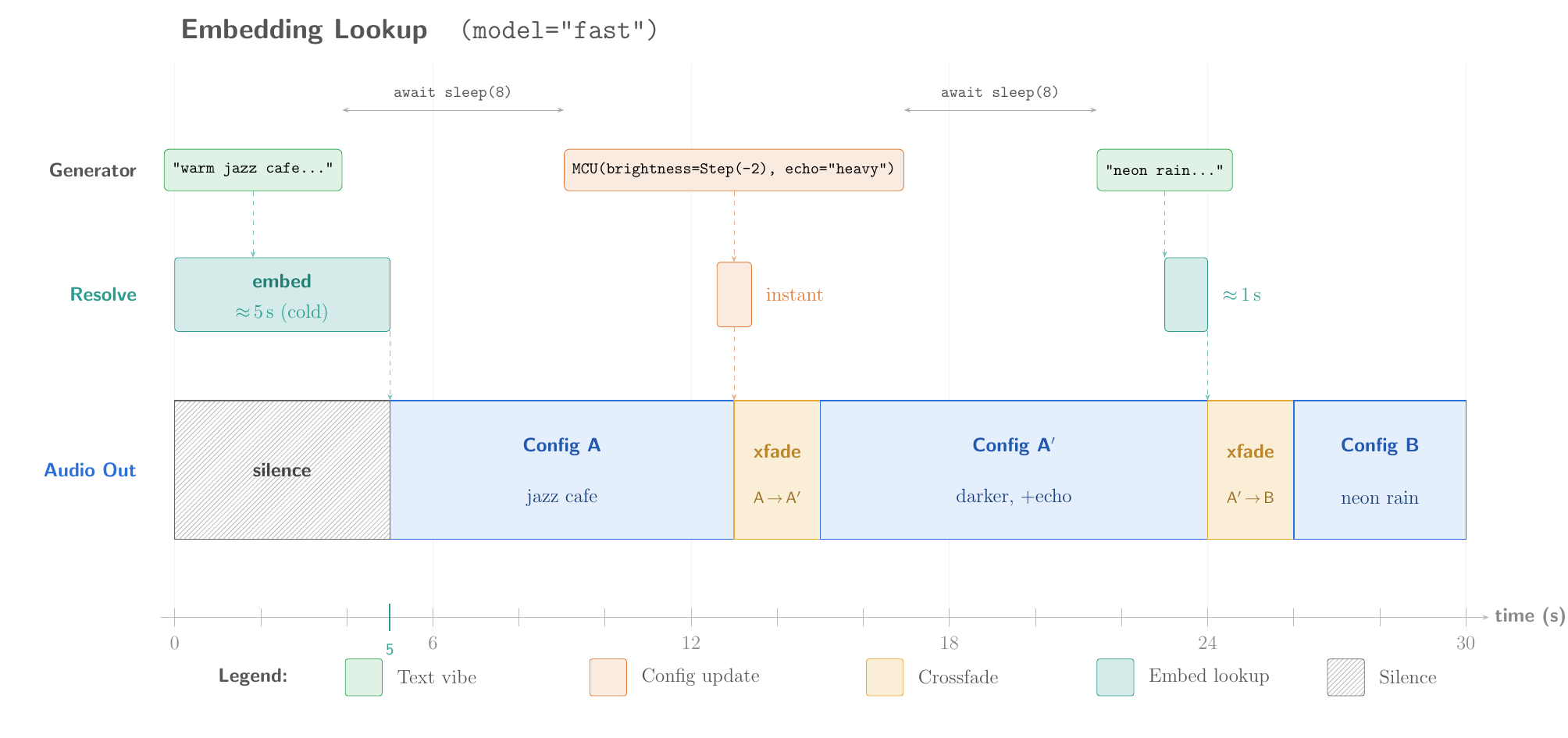}
  \caption{Embedding Lookup (\texttt{model="fast"}) backend. The first resolve includes model loading ({\raise.17ex\hbox{$\scriptstyle\sim$}}5\,s cold start); subsequent embed lookups resolve in {\raise.17ex\hbox{$\scriptstyle\sim$}}1\,s. Parameter updates (\texttt{MusicConfigUpdate}) apply instantly. Audio plays continuously after the first configuration resolves.}
  \Description{Timing diagram showing three horizontal swim lanes (Generator, Resolve, Audio Out) over 30 seconds. A hatched silence region spans 0--5s during cold start, then Config A plays from 5--13s, a 2s crossfade transitions to Config A-prime which plays from 15--24s, then another crossfade leads to Config B from 26--30s. Resolve boxes show a 5s cold embed lookup, an instant MCU update, and a 1s warm embed lookup.}
  \label{fig:timing-embed}
\end{figure*}

\begin{figure*}[t]
  \centering
  \includegraphics[width=\textwidth]{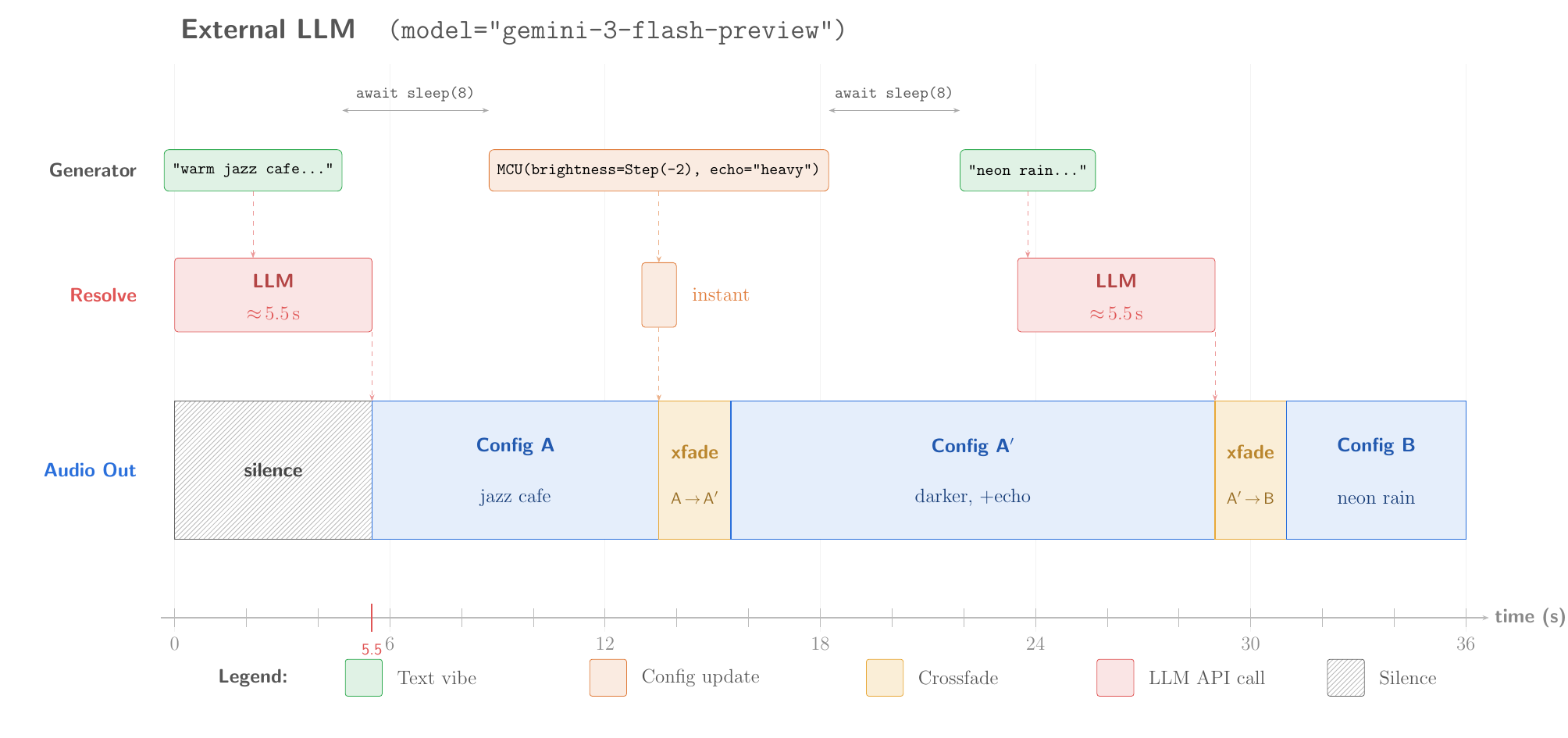}
  \caption{External LLM (\texttt{model="gemini-3-flash-preview"}) backend. Each text prompt triggers a {\raise.17ex\hbox{$\scriptstyle\sim$}}5.5\,s API call; the current configuration keeps playing uninterrupted while the new one resolves. The time axis extends to 36\,s for the same three instructions that the fast backend completes by 30\,s---this gap compounds with each text prompt.}
  \Description{Timing diagram showing three horizontal swim lanes over 36 seconds. Layout mirrors Figure 2 but with longer LLM resolve blocks of approximately 5.5 seconds each. Config A-prime plays for an extended period from 15.5s to 29s while the second LLM call resolves, demonstrating continuous audio despite backend latency.}
  \label{fig:timing-llm}
\end{figure*}

The sleep duration controls \emph{how long each configuration plays}, and new instructions can be yielded at any time---including while a previous resolution is still pending. Figures~\ref{fig:timing-embed} and~\ref{fig:timing-llm} show the timing relationship between the generator, resolver, and audio output for both backends. The SDK queues instructions and resolves them in order. Rapid yielding can accumulate pending instructions with slow backends; with the fast backend, rapid yields produce transitions too quick to sound musical. In practice, performers quickly learn to pace yields to the backend's resolution speed.

The SDK supports three backends. The \textbf{fast} mode (default) uses embedding retrieval---sub-second after warm-up, fully reliable, no network needed. The \textbf{external} mode routes to any LLM via the developer's API key, with seamless playback despite network latency; failed calls fall back to fast mode. A local \textbf{expressive} mode runs a 270M fine-tuned model on-device (experimental). A web-based UI demonstrates the SDK as a playable instrument: it displays each configuration's co-generated title and color palette alongside the parameter editor, with background particle effects responsive to playback state---creating an audiovisual performance surface.

\subsection{Configuration Schema as Mapping Design}

The schema contains 34~fields in five groups: global parameters~(8: \texttt{tempo}, \texttt{root}, \texttt{mode}, \texttt{brightness}, \texttt{space}, \texttt{density}, \texttt{motion}, \texttt{attack}), six orchestration layers (\texttt{bass}, \texttt{pad}, \texttt{melody}, \texttt{rhythm}, \texttt{texture}, \texttt{accent}---each selecting a style from a curated set), spatial/\allowbreak{}texture~(5: \texttt{stereo}, \texttt{depth}, \texttt{echo}, \texttt{human}, \texttt{grain}), melody generation~(10), and harmony~(5). All fields use categorical labels, designed for both human interpretability and reliable LLM generation---an LLM can produce \texttt{"bright"} far more reliably than a meaningful float like~\texttt{0.73}.

Eight of these fields use ordered labels that support relative stepping: a performer can issue \texttt{Step(+1)} or \texttt{Step(-1)} to move along the label sequence, clamping at boundaries. The remaining fields use style selectors, booleans, or bounded values accepting absolute assignments only. Each resolved configuration also includes co-generated metadata---a descriptive title, three color palettes, and a brief justification of the parameter choices.

The schema was iteratively refined through an AI-human design loop. Categorical labels make each level perceptually distinct and enable reliable LLM generation. Constrained style sets reduce discordant combinations. The design goal: \emph{most valid configurations should sound musically coherent.} The schema is biased toward coherence to support exploratory play---a decision in Magnusson's sense~\cite{magnusson2019sonic}, not a limitation.

\section{Retrieval-Based Configuration}

Given a coherent parameter space, the remaining challenge is mapping text to configurations quickly enough for live use. The default fast backend uses a retrieval map of LLM-generated configurations curated offline. We distilled {\raise.17ex\hbox{$\scriptstyle\sim$}}10,500 unique scene descriptions from the Common Pile~\cite{kandpal2025commonpile}, an openly licensed text corpus, prompted Gemini~3 Flash Preview to generate five candidate configurations per scene as valid schema instances, scored each rendered audio by measuring text--audio semantic similarity using LAION-CLAP~\cite{wu2023clap}, a contrastive language--audio model trained on diverse audio--text pairs, and selected the best, embedded each scene with a sentence transformer~\cite{reimers2019sentencebert}, and exported the map. At runtime, the nearest neighbor is returned---approximately one second on CPU.

\section{Design Observations}

The author used the instrument across multiple composition sessions. The live generator excels at gradual scenic evolution: yielding \texttt{"quiet midnight hush"}, waiting, then stepping \texttt{brightness} down and switching \texttt{rhythm} to \texttt{"heartbeat"} produces a coherent ambient environment that evolves naturally. This two-level control---text for macro-level scene changes, parameter stepping for micro-level refinement---emerged as the instrument's natural performance idiom. When using an external LLM backend, the 5--8\,s wait for a new scene to resolve is noticeable but not disruptive---the current soundscape keeps playing, and the crossfade when it arrives feels like a natural transition rather than a technical artifact. The instrument struggles with sharp stylistic pivots and cannot match timbral specificity for prompts like ``upright bass.''

Five lay participants each listened to four text-generated soundscapes paired with author-steered variations in randomized order, without being told which was which, and were asked to describe what they heard and whether changes felt intentional. All five independently described output as sounding musically intentional rather than random, offering informal support for the schema design goal. Limitations identified: no individual layer volume control, restricted stylistic range for culturally specific music, and occasional audio artifacts in dense configurations. These comments are treated as informal design feedback rather than evidence of general audience response: the sample was small, participants were lay listeners, and no statistical claims are made from these remarks.

A technical benchmark on 200 held-out prompts showed embedding retrieval scoring higher on text--audio semantic similarity than a random valid configuration, suggesting that retrieval adds semantic targeting beyond the schema's baseline coherence under this proxy. This metric is a system-reliability and alignment proxy, not a perceptual-quality measure. LAION-CLAP also informed retrieval-map construction, so retrieval has an expected advantage on this metric; human perceptual validation is planned.

\section{Discussion}

G1~(immediacy) and G3~(reproducibility) are met. G5~(continuity) is met---in a 200-prompt benchmark, the embedding backend resolved all prompts successfully ({\raise.17ex\hbox{$\scriptstyle\sim$}}0.3\,s config generation, {\raise.17ex\hbox{$\scriptstyle\sim$}}1.2\,s total with synthesis); the external LLM backend resolved 178/200 ({\raise.17ex\hbox{$\scriptstyle\sim$}}5.6\,s config generation, 89\% success rate), with failed calls falling back to the fast backend to maintain uninterrupted playback. G2~(controllability) is partially met---categorical labels provide precise steering, but retrieval coverage gaps cause occasional semantic mismatches. G4~(learnability) is informally supported but formally unevaluated.

The live generator reframes text-to-music from a one-shot generation into an \emph{ongoing stream}---decoupling resolution from emission generalizes beyond music to any latency-sensitive domain. The CPU-only, reproducible design responds to Clarke et~al.'s~\cite{clarke2025longevity} concerns about longevity, reproducibility, documentation, and the resource demands of deep generative models in NIME practice---the default backend requires no inference at runtime, amortizing LLM cost into a one-time dataset construction step. The open-source SDK, extensible schema, and shared retrieval dataset contribute to the NIME~2026 \emph{Communities} theme: the instrument's parameter space is a shared vocabulary that other developers can extend with new style sets, new retrieval entries, or new synthesizer backends---and the live generator architecture provides the runtime for any such extension to be performed live.

\paragraph{Musical outcomes and limitations.}
The strongest outputs are ambient, cinematic, and loop-adjacent soundscapes where texture and harmonic color matter more than authored melody. It is less successful for genre imitation or timbral specificity. This trade-off privileges predictable steering and live continuity over timbral range.

\section{Ethical Standards}

This work follows NIME ethical expectations. Informal listener feedback was solicited from consenting adult participants on coherence, controllability, and musical character; no personal data was collected, and the feedback is reported as informal design input, not a powered study. The system generates procedural audio rather than imitating recordings or performers; the retrieval dataset uses an openly licensed corpus~\cite{kandpal2025commonpile} and releases structured configurations and rendered examples rather than copyrighted music recordings. The CPU-first default backend amortizes LLM cost across all use, addressing NIME's guidelines on lightweight models and AI environmental impact. AI coding tools assisted with codebase and schema design, and grammatical correction; the open-source release includes an \texttt{AI\_DISCLOSURE.md} documenting their scope. A supplementary video demonstrates the live generator.

\bibliographystyle{ACM-Reference-Format}
\bibliography{references}

\appendix
\onecolumn
\raggedbottom

\hypersetup{colorlinks=true, linkcolor=blue!60!black, citecolor=blue!60!black, urlcolor=blue!60!black}

\lstset{
  basicstyle=\footnotesize\ttfamily,
  keywordstyle=\color{codeblue}\bfseries,
  commentstyle=\itshape\color{codegray},
  stringstyle=\color{codegreen},
  backgroundcolor=\color{backcolour},
  breaklines=true,
  frame=single,
  numbers=none,
  columns=fullflexible,
  aboveskip=8pt,
  belowskip=8pt,
  showstringspaces=false,
}
\setlist{nosep, leftmargin=*}
\setlength{\parindent}{0pt}
\setlength{\parskip}{6pt}

\section{Supplementary Material}
\label{sec:supp}

This appendix supplements the main paper with additional technical detail and artifact links. The open-source SDK and demo are available at \url{https://github.com/prabal-rje/latentscore} and \url{https://latentscore.com}.

\noindent The subsections below cover: \S\ref{sec:schema}~the full configuration schema including co-generated metadata, \S\ref{sec:benchmark}~benchmark results across six named controllers, \S\ref{sec:examples}~example outputs with color palettes, \S\ref{sec:dataset-availability}~dataset and artifact availability, and \S\ref{sec:figures}~supplementary figures.

\subsection{Configuration Schema Reference}
\label{sec:schema}

Each output produced by the system is a JSON object with four top-level fields:

\begin{itemize}
  \item \texttt{thinking} --- a short generated rationale summarizing the model's parameter choices; used as metadata during dataset construction and omitted from the performer's direct control surface.
  \item \texttt{title} --- a short evocative name ($\leq$6 words) summarizing the vibe, displayed in the UI as a human-readable label for the current soundscape.
  \item \texttt{config} --- the 34-field synthesis configuration (Table~\ref{tab:schema}).
  \item \texttt{palettes} --- three color palettes, each containing five hex colors with size weights (\texttt{xs}--\texttt{xxl}). These drive the UI's visual feedback: background gradients, particle colors, and ambient lighting shift in real time as configurations change. The co-generation of audio parameters and visual palettes from a single text prompt makes the system an \emph{audiovisual} instrument.
\end{itemize}

The web-based UI displays the title, renders the active palette as a color-shifting background, and overlays particle effects responsive to playback state. Together with the text input and parameter steering buttons, this creates a playable audiovisual performance surface where text prompts produce coordinated shifts in sound, color, and motion.

\subsubsection*{Synthesis Configuration Fields}

The \texttt{config} object contains 34~fields in five groups. Eight fields marked $\star$ support ordered stepping: \texttt{Step(+1)} or \texttt{Step(-1)} adjustments that clamp at boundaries.

{\small
\begin{longtable}{@{}>{\raggedright\arraybackslash}p{3.8cm}p{1.5cm}p{8.6cm}@{}}
\caption{Full configuration schema. Fields marked $\star$ support relative \texttt{Step} adjustments.}
\label{tab:schema}\\
\toprule
\textbf{Field} & \textbf{Type} & \textbf{Allowed Values} \\
\midrule
\endfirsthead
\toprule
\textbf{Field} & \textbf{Type} & \textbf{Allowed Values} \\
\midrule
\endhead
\bottomrule
\endfoot
\multicolumn{3}{@{}l}{\textbf{Global Parameters (8)}} \\
\midrule
\texttt{tempo}\,$\star$ & Ord.\ label & \texttt{very\_slow} $|$ \texttt{slow} $|$ \texttt{medium} $|$ \texttt{fast} $|$ \texttt{very\_fast} \\
\texttt{root} & Enum & \texttt{c} $|$ \texttt{c\#} $|$ \texttt{d} $|$ \texttt{d\#} $|$ \texttt{e} $|$ \texttt{f} $|$ \texttt{f\#} $|$ \texttt{g} $|$ \texttt{g\#} $|$ \texttt{a} $|$ \texttt{a\#} $|$ \texttt{b} \\
\texttt{mode} & Enum & \texttt{major} $|$ \texttt{minor} $|$ \texttt{dorian} $|$ \texttt{mixolydian} \\
\texttt{brightness}$\,\star$ & Ord.\ label & \texttt{very\_dark} $|$ \texttt{dark} $|$ \texttt{medium} $|$ \texttt{bright} $|$ \texttt{very\_bright} \\
\texttt{space}\,$\star$ & Ord.\ label & \texttt{dry} $|$ \texttt{small} $|$ \texttt{medium} $|$ \texttt{large} $|$ \texttt{vast} \\
\texttt{density}\,$\star$ & Bnd.\ int & \texttt{2} $|$ \texttt{3} $|$ \texttt{4} $|$ \texttt{5} $|$ \texttt{6} \\
\texttt{motion}\,$\star$ & Ord.\ label & \texttt{static} $|$ \texttt{slow} $|$ \texttt{medium} $|$ \texttt{fast} $|$ \texttt{chaotic} \\
\texttt{attack} & Enum & \texttt{soft} $|$ \texttt{medium} $|$ \texttt{sharp} \\
\midrule
\multicolumn{3}{@{}l}{\textbf{Orchestration Layers (6)}} \\
\midrule
\texttt{bass} & Style sel. & \texttt{drone} $|$ \texttt{sustained} $|$ \texttt{pulsing} $|$ \texttt{walking} $|$ \texttt{fifth\_drone} $|$ \texttt{sub\_pulse} $|$ \texttt{octave} $|$ \texttt{arp\_bass} \\
\texttt{pad} & Style sel. & \texttt{warm\_slow} $|$ \texttt{dark\_sustained} $|$ \texttt{cinematic} $|$ \texttt{thin\_high} $|$ \texttt{ambient\_drift} $|$ \texttt{stacked\_fifths} $|$ \texttt{bright\_open} \\
\texttt{melody} & Style sel. & \texttt{procedural} $|$ \texttt{contemplative} $|$ \texttt{rising} $|$ \texttt{falling} $|$ \texttt{minimal} $|$ \texttt{ornamental} $|$ \texttt{arp\_melody} $|$ \texttt{contemplative\_minor} $|$ \texttt{call\_response} $|$ \texttt{heroic} \\
\texttt{rhythm} & Style sel. & \texttt{none} $|$ \texttt{minimal} $|$ \texttt{heartbeat} $|$ \texttt{soft\_four} $|$ \texttt{hats\_only} $|$ \texttt{electronic} $|$ \texttt{kit\_light} $|$ \texttt{kit\_medium} $|$ \texttt{military} $|$ \texttt{tabla\_essence} $|$ \texttt{brush} \\
\texttt{texture} & Style sel. & \texttt{none} $|$ \texttt{shimmer} $|$ \texttt{shimmer\_slow} $|$ \texttt{vinyl\_crackle} $|$ \texttt{breath} $|$ \texttt{stars} $|$ \texttt{glitch} $|$ \texttt{noise\_wash} $|$ \texttt{crystal} $|$ \texttt{pad\_whisper} \\
\texttt{accent} & Style sel. & \texttt{none} $|$ \texttt{bells} $|$ \texttt{pluck} $|$ \texttt{chime} $|$ \texttt{bells\_dense} $|$ \texttt{blip} $|$ \texttt{blip\_random} $|$ \texttt{brass\_hit} $|$ \texttt{wind} $|$ \texttt{arp\_accent} $|$ \texttt{piano\_note} \\
\midrule
\multicolumn{3}{@{}l}{\textbf{Spatial / Texture (5)}} \\
\midrule
\texttt{stereo}\,$\star$ & Ord.\ label & \texttt{mono} $|$ \texttt{narrow} $|$ \texttt{medium} $|$ \texttt{wide} $|$ \texttt{ultra\_wide} \\
\texttt{depth} & Boolean & \texttt{true} $|$ \texttt{false} \\
\texttt{echo}\,$\star$ & Ord.\ label & \texttt{none} $|$ \texttt{subtle} $|$ \texttt{medium} $|$ \texttt{heavy} $|$ \texttt{infinite} \\
\texttt{human}\,$\star$ & Ord.\ label & \texttt{robotic} $|$ \texttt{tight} $|$ \texttt{natural} $|$ \texttt{loose} $|$ \texttt{drunk} \\
\texttt{grain} & Enum & \texttt{clean} $|$ \texttt{warm} $|$ \texttt{gritty} \\
\midrule
\multicolumn{3}{@{}l}{\textbf{Melody Generation (10)}} \\
\midrule
\texttt{melody\_engine} & Enum & \texttt{pattern} $|$ \texttt{procedural} \\
\texttt{phrase\_len\_bars} & Bnd.\ int & \texttt{2} $|$ \texttt{4} $|$ \texttt{8} \\
\texttt{melody\_density} & Ord.\ label & \texttt{very\_sparse} $|$ \texttt{sparse} $|$ \texttt{medium} $|$ \texttt{busy} $|$ \texttt{very\_busy} \\
\texttt{syncopation} & Ord.\ label & \texttt{straight} $|$ \texttt{light} $|$ \texttt{medium} $|$ \texttt{heavy} \\
\texttt{swing} & Ord.\ label & \texttt{none} $|$ \texttt{light} $|$ \texttt{medium} $|$ \texttt{heavy} \\
\texttt{motif\_repeat\_prob} & Ord.\ label & \texttt{rare} $|$ \texttt{sometimes} $|$ \texttt{often} \\
\texttt{step\_bias} & Enum & \texttt{step} $|$ \texttt{balanced} $|$ \texttt{leapy} \\
\texttt{chromatic\_prob} & Ord.\ label & \texttt{none} $|$ \texttt{light} $|$ \texttt{medium} $|$ \texttt{heavy} \\
\texttt{register\_min\_oct} & Bnd.\ int & \texttt{1}\,--\,\texttt{8} \\
\texttt{register\_max\_oct} & Bnd.\ int & \texttt{1}\,--\,\texttt{8} (must exceed min) \\
\midrule
\multicolumn{3}{@{}l}{\textbf{Harmony (5)}} \\
\midrule
\texttt{cadence\_strength} & Ord.\ label & \texttt{weak} $|$ \texttt{medium} $|$ \texttt{strong} \\
\texttt{tension\_curve} & Enum & \texttt{arc} $|$ \texttt{ramp} $|$ \texttt{waves} \\
\texttt{harmony\_style} & Enum & \texttt{auto} $|$ \texttt{pop} $|$ \texttt{jazz} $|$ \texttt{cinematic} $|$ \texttt{ambient} \\
\texttt{chord\_change\_bars} & Ord.\ label & \texttt{very\_slow} $|$ \texttt{slow} $|$ \texttt{medium} $|$ \texttt{fast} \\
\texttt{chord\_extensions} & Enum & \texttt{triads} $|$ \texttt{sevenths} $|$ \texttt{lush} \\
\end{longtable}
}

\subsection{Benchmark Results}
\label{sec:benchmark}

We benchmarked six controllers on 200 held-out test prompts (TEST split), preventing data leakage into the embedding lookup map.

\subsubsection*{Controllers}

\begin{enumerate}
  \item \textbf{Random Baseline} --- uniformly random valid configuration (ignores prompt).
  \item \textbf{Base Untrained} --- \texttt{unsloth/gemma-3-270m-it} (Gemma~3, 270M parameters), no fine-tuning, constrained via Outlines grammar-guided decoding.
  \item \textbf{SFT Fine-tuned} --- same Gemma~3 270M model after supervised fine-tuning on TRAIN split with LoRA, constrained via Outlines.
  \item \textbf{Claude Opus~4.5} --- Anthropic's commercial API, structured output mode.
  \item \textbf{Gemini~3 Flash Preview} --- Google's commercial API (\texttt{gemini-3-flash-preview}), structured output mode.
  \item \textbf{Embedding Lookup} --- default fast backend; nearest-neighbor retrieval using \texttt{all-MiniLM-L6-v2} sentence embeddings (Reimers \& Gurevych, 2019).
\end{enumerate}

\subsubsection*{Results Summary}

\begin{table}[h]
\centering
\caption{Benchmark results across 200 held-out test prompts. CLAP = LAION-CLAP with discordance penalty.}
\label{tab:benchmark}
\begin{tabular}{@{}lccccc@{}}
\toprule
\textbf{Controller} & \textbf{CLAP}$\uparrow$ & \textbf{Schema} & \textbf{Config} & \textbf{Synth} & \textbf{Total} \\
 & \textbf{(mean)} & \textbf{Valid \%} & \textbf{Gen (s)} & \textbf{(s)} & \textbf{(s)} \\
\midrule
Random Baseline         & 0.139 & 100\% & $<$0.01 & 0.56 & 0.70 \\
Base Untrained (270M)   & 0.117 & 100\% & 59.1   & 0.38 & 59.7 \\
SFT Fine-tuned (270M)   & 0.140 & 91\%  & 99.5   & 0.47 & 100.2 \\
Claude Opus~4.5         & 0.137 & 100\% & 11.9   & 0.57 & 12.6 \\
Gemini~3 Flash Preview  & 0.158 & 89\%  & 5.6    & 0.73 & 6.5  \\
\textbf{Embed.\ Lookup} & \textbf{0.163} & \textbf{100\%} & \textbf{0.3} & \textbf{0.79} & \textbf{1.2} \\
\bottomrule
\end{tabular}
\end{table}

Key observations:

\begin{itemize}
  \item \textbf{Embedding Lookup achieves the highest CLAP (0.163) with lowest total latency ({\raise.17ex\hbox{$\scriptstyle\sim$}}1.2\,s) and 100\% validity.} This is the default backend for live performance.
  \item Gemini~3 Flash Preview produces competitive CLAP (0.158) on valid outputs but fails schema validation 11\% of the time; in live use, such failures trigger automatic fallback to the fast backend.
  \item Claude Opus~4.5 achieves 100\% validity but scores below random on CLAP (0.137 vs.\ 0.139) with {\raise.17ex\hbox{$\scriptstyle\sim$}}12.6\,s mean latency.
  \item The fine-tuned 270M model only marginally exceeds random on CLAP (0.140 vs.\ 0.139) at {\raise.17ex\hbox{$\scriptstyle\sim$}}100\,s mean generation time, suggesting mode collapse at this model scale.
  \item Synthesis time is configuration-dependent: richer configs from Embedding Lookup (0.79\,s) and Gemini (0.73\,s) render slower than simpler ones from Base (0.38\,s) and Random (0.56\,s).
\end{itemize}

\textbf{CLAP circularity.} The retrieval map was constructed with Best-of-5 CLAP-scored selection. The same metric is used for evaluation; the retrieval advantage is partly expected. Human perceptual evaluation is planned.

\subsection{Example Configurations}
\label{sec:examples}

Three complete outputs from the retrieval dataset, each showing the full structure: \texttt{thinking} (the LLM's justification of its choices), \texttt{title}, all 34 \texttt{config} fields, and three co-generated \texttt{palettes} with color swatches.

\medskip\noindent\textbf{``Warm, comforting recollection.''}\par\nopagebreak[3]

\textbf{Title:} \textit{Hazy Echoes of a Gentle Memory} \quad\textbf{CLAP:} 0.569 \quad\textbf{Split:} SFT-Train

\vspace{2pt}
\textbf{Thinking:}
{\small\itshape
``A warm, comforting recollection calls for slow tempo and soft dynamics---like something remembered through a haze. F major gives a gentle warmth. Dark brightness with large space creates the sense of a memory filling a room. Vinyl crackle as texture reinforces the nostalgic, analog quality. Chime accents suggest fleeting moments surfacing. Ambient drift pad with sustained bass provides the continuous, enveloping foundation. Sparse melody with stepwise motion and weak cadences lets the piece breathe without urgency. Lush chord extensions add harmonic richness without tension.''}

\begin{lstlisting}[language=json]
{
  "tempo": "slow", "root": "f", "mode": "major",
  "brightness": "dark", "space": "large", "density": 3,
  "bass": "sustained", "pad": "ambient_drift",
  "melody": "contemplative", "rhythm": "none",
  "texture": "vinyl_crackle", "accent": "chime",
  "motion": "slow", "attack": "soft",
  "stereo": "wide", "depth": true,
  "echo": "subtle", "human": "natural", "grain": "warm",
  "melody_engine": "procedural", "phrase_len_bars": 8,
  "melody_density": "sparse", "syncopation": "straight",
  "swing": "none", "motif_repeat_prob": "sometimes",
  "step_bias": "step", "chromatic_prob": "none",
  "cadence_strength": "weak",
  "register_min_oct": 3, "register_max_oct": 5,
  "tension_curve": "arc", "harmony_style": "ambient",
  "chord_change_bars": "slow", "chord_extensions": "lush"
}
\end{lstlisting}

\textbf{Palette 1:}\enspace
\colorbox[HTML]{f4e1d2}{\phantom{XX}}\,{\footnotesize\texttt{\#f4e1d2}\,(xxl)}\enspace
\colorbox[HTML]{e9c46a}{\phantom{XX}}\,{\footnotesize\texttt{\#e9c46a}\,(xl)}\enspace
\colorbox[HTML]{f4a261}{\phantom{XX}}\,{\footnotesize\texttt{\#f4a261}\,(lg)}\enspace
\colorbox[HTML]{e76f51}{\phantom{XX}}\,{\footnotesize\texttt{\#e76f51}\,(md)}\enspace
\colorbox[HTML]{264653}{\color{white}\phantom{XX}}\,{\footnotesize\texttt{\#264653}\,(sm)}

\textbf{Palette 2:}\enspace
\colorbox[HTML]{fdfcf0}{\phantom{XX}}\,{\footnotesize\texttt{\#fdfcf0}\,(xxl)}\enspace
\colorbox[HTML]{faedcd}{\phantom{XX}}\,{\footnotesize\texttt{\#faedcd}\,(xl)}\enspace
\colorbox[HTML]{d4a373}{\phantom{XX}}\,{\footnotesize\texttt{\#d4a373}\,(lg)}\enspace
\colorbox[HTML]{ccd5ae}{\phantom{XX}}\,{\footnotesize\texttt{\#ccd5ae}\,(md)}\enspace
\colorbox[HTML]{e9edc9}{\phantom{XX}}\,{\footnotesize\texttt{\#e9edc9}\,(sm)}

\textbf{Palette 3:}\enspace
\colorbox[HTML]{efd3d7}{\phantom{XX}}\,{\footnotesize\texttt{\#efd3d7}\,(xxl)}\enspace
\colorbox[HTML]{feeafa}{\phantom{XX}}\,{\footnotesize\texttt{\#feeafa}\,(xl)}\enspace
\colorbox[HTML]{dee2ff}{\phantom{XX}}\,{\footnotesize\texttt{\#dee2ff}\,(lg)}\enspace
\colorbox[HTML]{8e9aaf}{\color{white}\phantom{XX}}\,{\footnotesize\texttt{\#8e9aaf}\,(md)}\enspace
\colorbox[HTML]{cbc0d3}{\phantom{XX}}\,{\footnotesize\texttt{\#cbc0d3}\,(sm)}

\medskip\noindent\textbf{``Need for continual self-examination.''}\par\nopagebreak[3]

\textbf{Title:} \textit{Infinite Internal Reflection} \quad\textbf{CLAP:} 0.556 \quad\textbf{Split:} SFT-Train

\vspace{2pt}
\textbf{Thinking:}
{\small\itshape
``Self-examination is inward-looking, contemplative, cyclical. Dorian mode in D gives a modal ambiguity---not quite sad, not resolved. A drone bass anchors the introspection; ambient drift pad creates a sense of floating within one's own thoughts. No rhythm---this is internal, not measured by external time. Pad whisper texture adds subliminal movement beneath the surface. Piano note accents surface like individual thoughts emerging. Slow motion, soft attack, medium stereo---contained but not claustrophobic. Waves tension curve for the recurring nature of self-questioning. Often-repeating motifs suggest obsessive return to the same themes.''}

\begin{lstlisting}[language=json]
{
  "tempo": "slow", "root": "d", "mode": "dorian",
  "brightness": "dark", "space": "large", "density": 3,
  "bass": "drone", "pad": "ambient_drift",
  "melody": "contemplative", "rhythm": "none",
  "texture": "pad_whisper", "accent": "piano_note",
  "motion": "slow", "attack": "soft",
  "stereo": "medium", "depth": true,
  "echo": "subtle", "human": "natural", "grain": "warm",
  "melody_engine": "procedural", "phrase_len_bars": 8,
  "melody_density": "sparse", "syncopation": "straight",
  "swing": "none", "motif_repeat_prob": "often",
  "step_bias": "step", "chromatic_prob": "light",
  "cadence_strength": "weak",
  "register_min_oct": 2, "register_max_oct": 4,
  "tension_curve": "waves", "harmony_style": "ambient",
  "chord_change_bars": "slow", "chord_extensions": "lush"
}
\end{lstlisting}

\textbf{Palette 1:}\enspace
\colorbox[HTML]{1a1a2e}{\color{white}\phantom{XX}}\,{\footnotesize\texttt{\#1a1a2e}\,(xxl)}\enspace
\colorbox[HTML]{16213e}{\color{white}\phantom{XX}}\,{\footnotesize\texttt{\#16213e}\,(xl)}\enspace
\colorbox[HTML]{0f3460}{\color{white}\phantom{XX}}\,{\footnotesize\texttt{\#0f3460}\,(lg)}\enspace
\colorbox[HTML]{533483}{\color{white}\phantom{XX}}\,{\footnotesize\texttt{\#533483}\,(md)}\enspace
\colorbox[HTML]{e94560}{\color{white}\phantom{XX}}\,{\footnotesize\texttt{\#e94560}\,(sm)}

\textbf{Palette 2:}\enspace
\colorbox[HTML]{2c3e50}{\color{white}\phantom{XX}}\,{\footnotesize\texttt{\#2c3e50}\,(xxl)}\enspace
\colorbox[HTML]{34495e}{\color{white}\phantom{XX}}\,{\footnotesize\texttt{\#34495e}\,(xl)}\enspace
\colorbox[HTML]{7f8c8d}{\color{white}\phantom{XX}}\,{\footnotesize\texttt{\#7f8c8d}\,(lg)}\enspace
\colorbox[HTML]{95a5a6}{\phantom{XX}}\,{\footnotesize\texttt{\#95a5a6}\,(md)}\enspace
\colorbox[HTML]{bdc3c7}{\phantom{XX}}\,{\footnotesize\texttt{\#bdc3c7}\,(sm)}

\textbf{Palette 3:}\enspace
\colorbox[HTML]{000000}{\color{white}\phantom{XX}}\,{\footnotesize\texttt{\#000000}\,(xxl)}\enspace
\colorbox[HTML]{141414}{\color{white}\phantom{XX}}\,{\footnotesize\texttt{\#141414}\,(xl)}\enspace
\colorbox[HTML]{1f1f1f}{\color{white}\phantom{XX}}\,{\footnotesize\texttt{\#1f1f1f}\,(lg)}\enspace
\colorbox[HTML]{3d3d3d}{\color{white}\phantom{XX}}\,{\footnotesize\texttt{\#3d3d3d}\,(md)}\enspace
\colorbox[HTML]{ffffff}{\phantom{XX}}\,{\footnotesize\texttt{\#ffffff}\,(sm)}

\medskip\noindent\textbf{``The mood is anticipatory.''}\par\nopagebreak[3]

\textbf{Title:} \textit{The Growing Momentum of Expectation} \quad\textbf{CLAP:} 0.554 \quad\textbf{Split:} SFT-Train

\vspace{2pt}
\textbf{Thinking:}
{\small\itshape
``Anticipation means something is building. Medium tempo with slow motion creates a measured but unstoppable forward movement. Dorian mode for tension without darkness. Bright brightness---anticipation is alert, not gloomy. Sub-pulse bass provides the physical heartbeat of waiting. Heartbeat rhythm reinforces this literally. Rising melody enacts the building expectation; ramp tension curve ensures it never resolves too soon. Cinematic pad and cinematic harmony style for the grand-scale, almost filmic sense of approaching something. Sparse melody density with light syncopation---not rushed, but not metronomic. Wide stereo and large space for the sense of something vast about to happen.''}

\begin{lstlisting}[language=json]
{
  "tempo": "medium", "root": "d", "mode": "dorian",
  "brightness": "bright", "space": "large", "density": 4,
  "bass": "sub_pulse", "pad": "cinematic",
  "melody": "rising", "rhythm": "heartbeat",
  "texture": "shimmer_slow", "accent": "blip",
  "motion": "slow", "attack": "medium",
  "stereo": "wide", "depth": true,
  "echo": "subtle", "human": "tight", "grain": "clean",
  "melody_engine": "procedural", "phrase_len_bars": 4,
  "melody_density": "sparse", "syncopation": "light",
  "swing": "none", "motif_repeat_prob": "sometimes",
  "step_bias": "balanced", "chromatic_prob": "light",
  "cadence_strength": "medium",
  "register_min_oct": 3, "register_max_oct": 6,
  "tension_curve": "ramp", "harmony_style": "cinematic",
  "chord_change_bars": "slow", "chord_extensions": "lush"
}
\end{lstlisting}

\textbf{Palette 1:}\enspace
\colorbox[HTML]{2c3e50}{\color{white}\phantom{XX}}\,{\footnotesize\texttt{\#2c3e50}\,(xxl)}\enspace
\colorbox[HTML]{34495e}{\color{white}\phantom{XX}}\,{\footnotesize\texttt{\#34495e}\,(xl)}\enspace
\colorbox[HTML]{7f8c8d}{\color{white}\phantom{XX}}\,{\footnotesize\texttt{\#7f8c8d}\,(lg)}\enspace
\colorbox[HTML]{bdc3c7}{\phantom{XX}}\,{\footnotesize\texttt{\#bdc3c7}\,(md)}\enspace
\colorbox[HTML]{ecf0f1}{\phantom{XX}}\,{\footnotesize\texttt{\#ecf0f1}\,(sm)}

\textbf{Palette 2:}\enspace
\colorbox[HTML]{1a1a2e}{\color{white}\phantom{XX}}\,{\footnotesize\texttt{\#1a1a2e}\,(xxl)}\enspace
\colorbox[HTML]{16213e}{\color{white}\phantom{XX}}\,{\footnotesize\texttt{\#16213e}\,(xl)}\enspace
\colorbox[HTML]{0f3460}{\color{white}\phantom{XX}}\,{\footnotesize\texttt{\#0f3460}\,(lg)}\enspace
\colorbox[HTML]{e94560}{\color{white}\phantom{XX}}\,{\footnotesize\texttt{\#e94560}\,(md)}\enspace
\colorbox[HTML]{ffffff}{\phantom{XX}}\,{\footnotesize\texttt{\#ffffff}\,(sm)}

\textbf{Palette 3:}\enspace
\colorbox[HTML]{000000}{\color{white}\phantom{XX}}\,{\footnotesize\texttt{\#000000}\,(xxl)}\enspace
\colorbox[HTML]{14213d}{\color{white}\phantom{XX}}\,{\footnotesize\texttt{\#14213d}\,(xl)}\enspace
\colorbox[HTML]{fca311}{\phantom{XX}}\,{\footnotesize\texttt{\#fca311}\,(lg)}\enspace
\colorbox[HTML]{e5e5e5}{\phantom{XX}}\,{\footnotesize\texttt{\#e5e5e5}\,(md)}\enspace
\colorbox[HTML]{ffffff}{\phantom{XX}}\,{\footnotesize\texttt{\#ffffff}\,(sm)}

\vspace{6pt}

Note how palette color temperatures correlate with audio parameters: warm/nostalgic vibes produce cream and amber tones; introspective vibes produce deep navy, indigo, and near-black; anticipatory vibes produce cool grays with accent pops. The \texttt{thinking} field is a short generated rationale captured during dataset construction; it is retained as metadata about parameter choices and is not part of the performer's control surface.

\subsection{Dataset and Artifact Availability}
\label{sec:dataset-availability}

The dataset and benchmark artifacts are archived through public Hugging Face repositories; the repository file listings provide the authoritative artifact record rather than the automatically generated dataset preview.

\begin{itemize}
  \item Dataset artifacts: \url{https://huggingface.co/datasets/guprab/latentscore-data}
  \item Benchmark artifacts and audio: \url{https://huggingface.co/datasets/guprab/latentscore-clap-benchmark}
  \item Model artifacts: \url{https://huggingface.co/guprab/latentscore-gemma3-270m-v5-merged}
\end{itemize}

\paragraph{Availability.}
The SDK, retrieval-map dataset artifacts, benchmark outputs, model artifacts, and demo materials are released through the public repository and the Hugging Face file listings linked above. Code and samples are available at \url{https://github.com/prabal-rje/latentscore}.

\subsection{Supplementary Figures}
\label{sec:figures}

\begin{figure}[H]
\centering
\includegraphics[width=\textwidth]{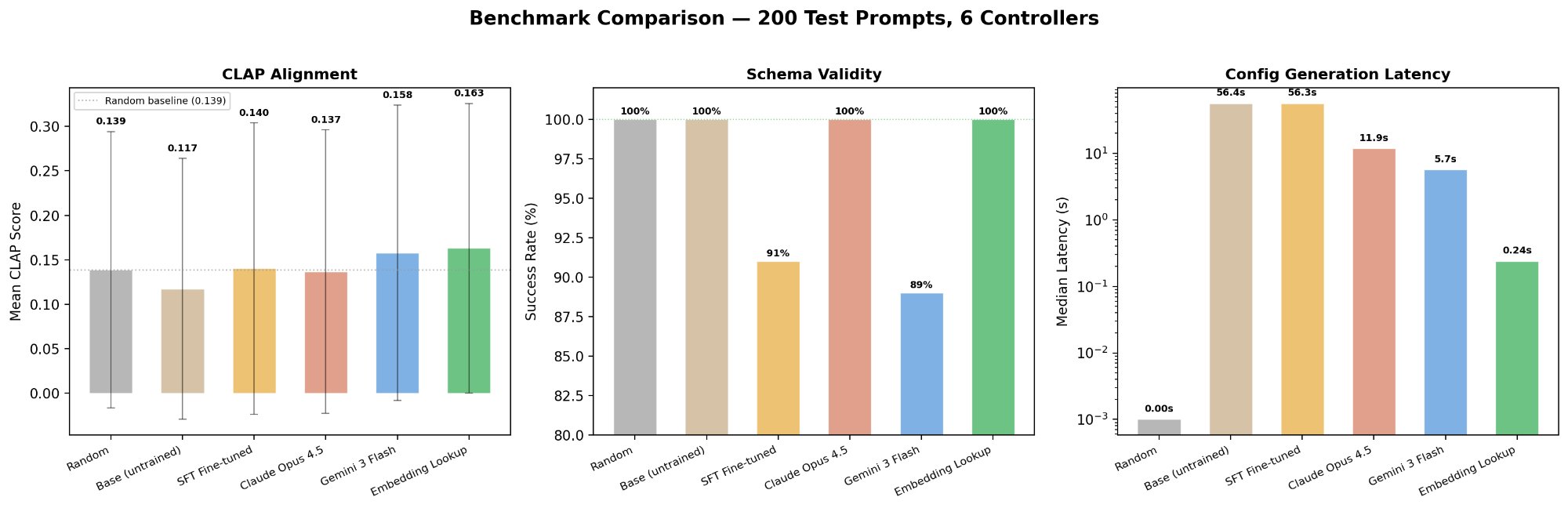}
\caption{Benchmark comparison across 200 test prompts and 6 controllers. \textnormal{\textbf{Left:}} CLAP alignment scores with standard deviation bars and random baseline reference line. \textnormal{\textbf{Center:}} Schema validity rates. \textnormal{\textbf{Right:}} Config generation latency on log scale, spanning three orders of magnitude from embedding retrieval (0.24\,s) to local 270M models ({\raise.17ex\hbox{$\scriptstyle\sim$}}56\,s).}
\label{fig:benchmark}
\end{figure}

\begin{figure}[H]
\centering
\includegraphics[width=\textwidth]{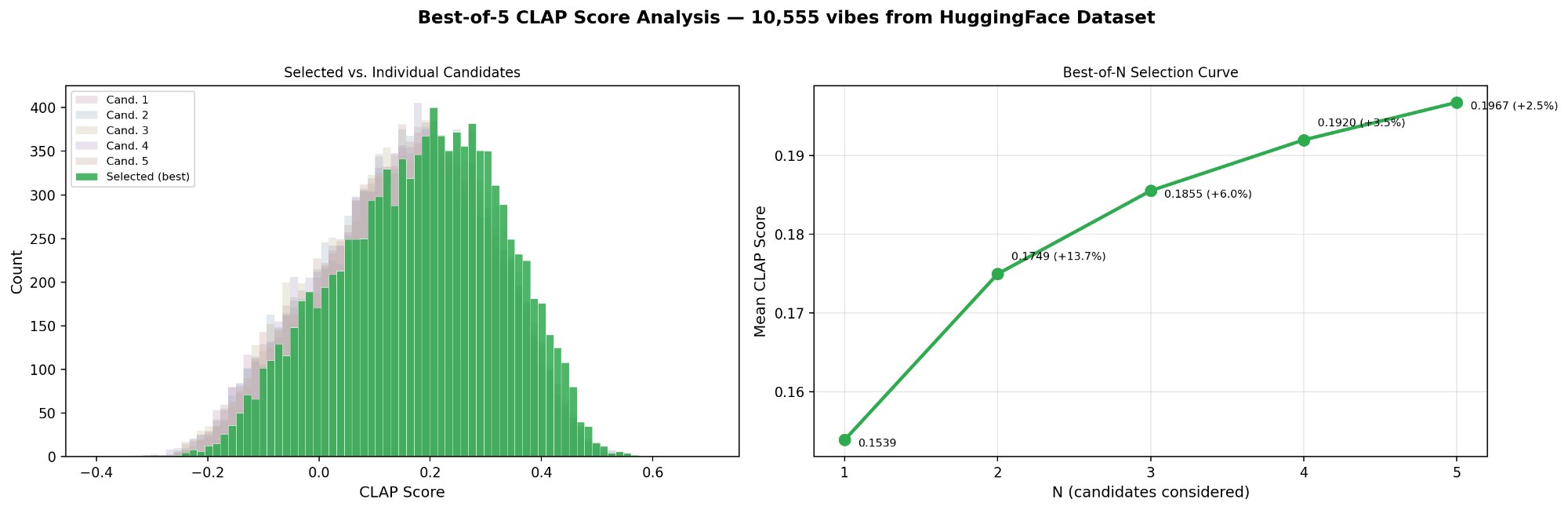}
\caption{Best-of-5 CLAP score analysis across 10,555 vibes in the dataset. \textnormal{\textbf{Left:}} Overlaid histograms of individual candidate scores (muted) vs.\ the selected best (green), showing the selected distribution shifted right. \textnormal{\textbf{Right:}} Best-of-N selection curve with diminishing returns: N\,=\,1 (0.154) to N\,=\,5 (0.197, +28\% cumulative).}
\label{fig:bestofn}
\end{figure}

\begin{figure}[H]
\centering
\includegraphics[width=\textwidth]{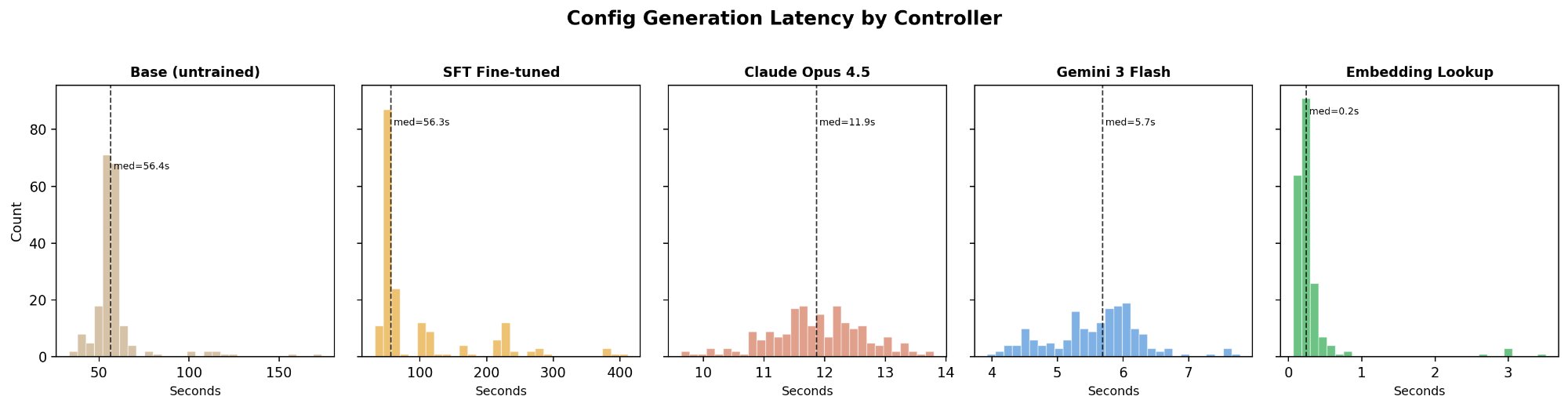}
\caption{Config generation latency histograms by controller. Gemma~3 270M models (Base and SFT, on H100 GPU) cluster around 56\,s with a long tail for SFT. Claude Opus~4.5 clusters tightly at 11.9\,s. Gemini~3 Flash Preview at 5.7\,s. Embedding Lookup at 0.2\,s median.}
\label{fig:config-latency}
\end{figure}

\begin{figure}[H]
\centering
\includegraphics[width=\textwidth]{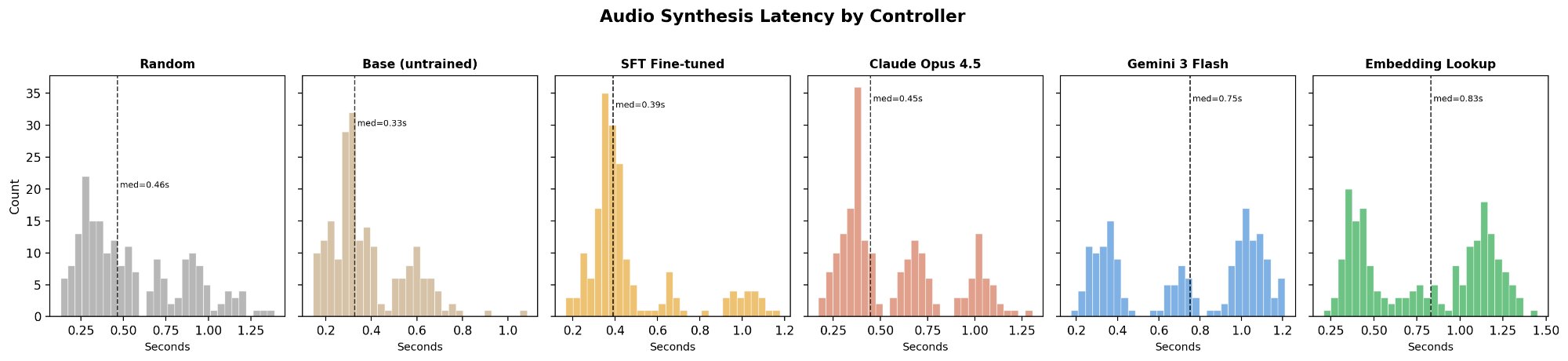}
\caption{Audio synthesis latency histograms (CPU procedural synthesizer, 30\,s renders). Synthesis time is configuration-dependent: richer configurations from Embedding Lookup (0.83\,s) and Gemini (0.75\,s) render slower than simpler ones from Base untrained (0.33\,s) and Random (0.46\,s).}
\label{fig:synth-latency}
\end{figure}

\vspace{12pt}
\noindent\rule{\textwidth}{0.4pt}

\small
References: Reimers \& Gurevych (2019), ``Sentence-BERT,'' EMNLP.\@ LAION-CLAP: Wu et al.\ (2023), ICASSP.\@ Common Pile: Kandpal et al.\ (2025). Gemma~3: Google (2025). Outlines: Willard \& Louf (2023). Full references appear in the bibliography above.

\end{document}